\documentclass{article}
\usepackage{spconf,amsmath,graphicx}
\usepackage{algorithm}
\usepackage{algorithmicx}
\usepackage{algpseudocode}
\usepackage{tabularx}
\usepackage{multirow}
\usepackage{epstopdf}
\usepackage{url}
\usepackage{etoolbox}

\ninept
\title{Audio Content based Geotagging in Multimedia}
\name{Anurag Kumar\textsuperscript{*}, Benjamin Elizalde\textsuperscript{+}, Bhiksha Raj\textsuperscript{*}}
\address{School of Computer Science\textsuperscript{*}, Electrical and Computer Engineering \textsuperscript{+} \\ Carnegie Mellon University\\Pittsburgh, PA, USA - 15213 \\ {\small \tt alnu@andrew.cmu.edu, bmartin1@andrew.cmu.edu, bhiksha@cs.cmu.edu} }
\begin{document}
%
\maketitle
\begin{abstract}
In this paper we propose methods to extract geographically relevant information in a multimedia recording using its audio. Our method primarily is based on the fact that urban acoustic environment consists of a variety of sounds. Hence, location information can be inferred from the composition of sound events/classes present in the audio. More specifically, we adopt matrix factorization techniques to obtain semantic content of recording in terms of different sound classes. These semantic information are then combined to identify the location of recording.
\end{abstract}
\vspace{-0.10in}
\section{Introduction}
\label{sec:intro}
\vspace{-0.1in}
Extracting information from multimedia recordings has received lot of attention due to the growing multimedia content on the web. A particularly interesting problem is extraction of geo-locations or information relevant to geographical locations. This process of providing geographical identity information is usually termed as Geotagging \cite{luo2011geotagging} and is gaining importance due its role in several applications. It is useful not only in location based services and recommender systems \cite{bao2012location} \cite{bao2015recommendations}\cite{majid2013context} but also in general cataloguing, organization, search and retrieval of multimedia content on the web. Location specific information also allows a user to put his/her multimedia content into a social context, since it is human nature to associate with geographical identity of any material. A nice survey on different aspects of geotagging in multimedia is provided in \cite{luo2011geotagging}.

Although, there are applications which allows users to add geographical information in their photos and videos, a larger portion of multimedia content on the web is without any geographical identity. In these cases geotags needs to be inferred from the multimedia content and the associated metadata. This problem of geotagging or location identification also features as the Placing Tasks in yearly MediaEval \cite{medeval} tasks. The goal of Placing Tasks \cite{choi2014placing} in MediaEval is to develop systems which can predict places in videos based on different modalities of multimedia such as images, audio, text etc. An important aspect of location prediction systems is the granularity at which location needs to be predicted. The Placing Task recognizes a wide range of location hierarchy, starting from neighbourhoods and going upto continents. In this work we are particularly interested in obtaining city-level geographical tags which is clearly one of the most important level of location specification for any data. City-level information is easily relatable and is well suited to location based services and recommender systems. 

Most of the current works on geotagging focus on using visual/image component of multimedia and the associated text in the multimedia (\cite{trevisiol2013retrieving} \cite{luo2011geotagging}\cite{song2012web}\cite{kelm2013novel} to cite a few). The audio component has largely been ignored and there is little work on predicting locations based on audio content of the multimedia. However, authors in \cite{choi2013human} argue that there are cases where audio content might be extremely helpful in identifying location. For example, speech based cues can aid in recognizing location. Moreover, factors such as urban soundscapes and locations acoustic environment can also help in location identification.  Very few works have looked into audio based location identification in multimedia recordings \cite{lei2012multimodal} \cite{sevillano2012audio}. The approaches proposed in these works have been simplistic relying mainly on basic low level acoustic features. One way is to use well known basic acoustic features such as Mel-Cepstra Coefficient (MFCC), Gammatone filter features directly for classification purposes. In other case audio-clip level features such \emph{GMM - Supervectors} or \emph{Bag Of Audio Words} (BoAW) histograms are first obtained and then classifiers are trained on these features. 

In this work we show that geotagging using \emph{only audio} component of multimedia can be done with reasonably good success rate by capturing the semantic content in the audio. Our primary assertion is that the semantic content of an audio recording in terms of different acoustic events can help in predicting locations. We argue that soundtracks of different cities are composed of a set of acoustic events. If we can somehow capture the composition of audio in terms of these acoustic events then they can be used to train machine learning algorithms for geotagging purposes. We start with a set of base acoustic events or sound classes and then apply methods based on matrix factorization to find the composition of soundtracks in terms of these acoustic events. Once the weights corresponding to each base sound class have been obtained, we build higher level feature using these weights which are further used to obtain kernel representations. The kernels corresponding to each basis sound are then combined to finally train Support Vector Machines for predicting location identification of the recording. 

The rest of the paper is organized as follows. In Section 2 we describe our proposed framework for audio based geotagging. In Section 3 we present our experiments and results. In Section 4 we  discuss scalability of our proposed method and also give concluding remarks.
\vspace{-0.1in}
\section{Audio Based Geotagging}
\label{sec:format}
\vspace{-0.1in}
Audio based geotagging in multimedia can be performed by exploiting audio content in several ways. One can possibly try to use automatic speech recognition (ASR) to exploit the speech information present in audio. For example, speech might contain words or sentences which uniquely identifies a place, \emph{I am near Eiffel Tower} clearly gives away the location as Paris, with high probability, irrespective of presence or absence of any other cues. Other details such as language used, mention of landmarks etc. in speech can also help in audio based geotagging. 

\subsection{Audio Semantic Content based Geotagging}
In this work we take a more generic approach where we try to capture semantic content of audio through occurrence of different meaningful sound events and scenes in the recording. We argue that it should be possible to train machines to capture identity of a location by capturing the composition of audio recordings in terms of human recognizable sound events. This idea can be related to and is in fact backed by urban soundscapes works \cite{brown2011towards} \cite{payne2009research}. Based on this idea of location identification through semantic content of audio, we try to answer two important questions. \emph{First}, how to \emph{mathematically} capture the composition of audio recordings and \emph{Second}, how to use the information about semantic content of the recording for training classifiers which can predict identity of location. We provide our answers for each of these questions one by one. It is worth noting that this overall framework is different from audio events recognition works. Our goal is not to identify acoustic events but to find the composition of acoustic events in a way which can further be used to obtain geographical locations. 

Let $E=\{E_1,E_2,..E_L\}$ be the set of base acoustic events or sound classes whose composition is to be captured in an audio recording. Let us assume that each of these sound classes can be characterized by a basis matrix $M_l$. For a given sound class $E_l$ the column vectors of its basis matrix $M_l$ essentially spans the space of sound class $E_l$. Mathematically, this span is in space of some acoustic feature (\emph{e.g} MFCC) used to characterize audio recordings and over which the basis matrices have been learned. How we obtain $M_l$ is discussed later. Any given soundtrack or audio recording is then decomposed with respect to the sound class $E_l$ as 
\begin{equation}
\label{eq:bsc}
X \approx M_lW_l^T
\end{equation}
where $X$ is a $d \times n $ dimensional representation of the audio recording using acoustic features such as MFCC. For MFCCs, this implies each column of $X$ is $d$ dimensional mel-frequency cepstral coefficients and $n$ is the total number of frames in the audio recording. The sound basis matrices $M_l$ are $d \times k$ dimensional where $k$ represents the number of basis vectors in $M_l$. In principle $k$ can vary with each sound class, however, for sake of convenience we assume it is same for all $E_l$, for $l=1 \,\,to\,\, L$. 

Equation 1 defines the relationship between the soundtrack and its composition in terms of sound classes. The weight matrix $W_l$ captures how the sound class $E_l$ is present in the recording. It is representative of the distribution of sound class $E_l$ through out the duration of the recording. Hencer, obtaining $W_l$ for each $l$ provides us information about the structural composition of the audio in terms of sound classes in $E$. These $W_l$ can then be used for differentiating locations. Thus, the first problem we need to address is to learn $M_l$ for each $E_l$ and then using it to compute $W_l$ for any given recording.
\subsection{Learning $\mathbf{M_l}$ and $\mathbf{W_l}$ using semi-NMF}
Let us assume that for a given sound class $E_l$ we have a collection of $N_l$ audio recordings belonging to class $E_l$ only. We parametrize each of these recordings through some acoustic features. In this work we use MFCC features augmented by delta and acceleration coefficients (denoted by MFCA) as basic acoustic features. These acoustic features are represented by $d \times n_i$ dimensional matrix $X_{E_l}^i$ for the $i^{th}$ recording. $d$ is dimensionality of acoustic features and each column represents acoustic features for a frame.  The basic features of all recordings are collected into one single matrix $X_{E_l} = [X_{E_l}^i,..X_{E_l}^N]$, to get a large collective sample of acoustic features for sound class $E_l$. Clearly, $X_{E_l}$ has $d$ rows and let $T$ be the number of columns in this matrix. 

To obtain the basis matrix $M_l$ for $E_l$ we employ matrix factorization techniques. More specifically, we use Non-Negative matrix factorization (NMF) like method proposed in \cite{ding2010convex}. \cite{ding2010convex} proposed two matrix factorization methods named \emph{semi-NMF} and \emph{convex-NMF} which are like NMF but do not require the matrix data to be non-negative. This is important in our case, since employing classical NMF \cite{lee2001algorithms} algorithms would require our basic acoustic feature to be non-negative. This can be highly restrictive given the challenging task at hand. Even though we employ MFCCs as acoustic features, our proposed general framework based on semi-NMF can be used with other features as well. Moreover, semi-NMF offers other interesting properties such as its \emph{interpretation} in terms of \emph{K-means clustering}. One of our higher level features is based on this interpretation of semi-NMF. convex-NMF did not yield desirable results and hence we do not discuss it in this paper. 

semi-NMF considers factorization of a matrix, $X_{E_l}$ as $X_{E_l} \approx M_lW^T$. For factorization number of basis vectors $k$ in $M_l$ is fixed to a value less than $min(d,T)$. semi-NMF does not impose any restriction on $M_l$, that is its element can have any sign. The weight matrix $W$ on the other hand is restricted to be non-negative. The objective is to minimize $||X_{E_l} - M_lW^T || ^2$. Assuming that $M_l$ and $W$  have been initialized, $M_l$ and $W_l$ are updated iteratively in the following way.  In each step of iteration, 
\begin{align}
\bullet \text{Fix} & \,\,W, \text{update}\,\, M_l \,\, as,\,\,M_l = X_{E_l}W(W^TW)^{-1}\\
\bullet \text{Fix} & \,\, M_l, \text{update} \,\,W, \,\, \resizebox{0.55\columnwidth}{!}{$ W_{rs} = W_{rs} \sqrt{\frac{(X^T_{E_l}M_l)^+_{rs} + [W(M_l^TM_l)^-]_{rs}}{(X_{E_l}^TM_l)^-_{rs} + [W(M_l^TM_l)^+]_{rs}}}$} \label{eq:wtup}
\end{align}

The process is iterated till error drops below certain tolerance. The $+$ and $-$ sign represents positive and negative parts of a matrix obtained as $Z^+_{rs} = (|Z_{rs}| + Z_{rs})/2$ and $Z^-_{rs} = (|Z_{rs}| - Z_{rs})/2$. Theoretical guarantees on convergence of semi-NMF and other interesting properties such as invariance with respect to scaling can be found in original paper. One interesting aspect of semi-NMF described by authors is its analysis in terms of K-means clustering algorithm. The objective function $||X-MW^T||^2$ can be related to K-Means objective function with $M_l$ representing the $k$ cluster centers. Hence, the basis matrix $M_l$ also represents centers of clusters. We exploit this interpretation in the next phase of our approach. The initialization of $M_l$ and $W_l$ is done as per the procedure described in \cite{ding2010convex}.  

Once $M_l$ have been learned for each $E_l$, we can easily obtain $W_l$ for any given audio recording $X$ by fixing $M_l$ and then applying Eq \ref{eq:wtup} for $X$ for several iterations. For a given $X$, $W_l$ contains information about $E_l$ in $X$. With K-Means interpretation of semi-NMF, the non-negative weight matrix $W_l$ can be interpreted as containing soft assignment posteriors to each cluster for all frames in $X$.
\subsection{Discriminative Learning using $W_l$}
We treat the problem of location prediction as a retrieval problem where we want to retrieve most relevant recordings belonging to a certain location (city). Put more formally, we train binary classifiers for each location to retrieve the most relevant recordings belonging to the concerned location. Let us assume that we are concerned with a particular city $C$ and the set $S=\{s_i,  i=1 \,\,to\,\,N\}$ is the set of available training audio recordings. The labels of the recordings are represented by $y_i \in \{-1,1\}$ with $y_i=1$ if $s_i$ belongs to $C$, otherwise $y_i=-1$. $X_i$ ($d \times n_i$) denotes the MFCA representation of $s_i$. For each $X_i$ weight composition matrices $W_i^l$ are obtained with respect to all sound events $E_l$ in $E$. $W_i^l$ captures distribution of sound event $E_l$ in $X_i$ and we propose $2$ histogram based representations to characterize this distribution.
\subsubsection{Direct characterization of $W_l$ as posterior }
\label{ssec:postr}
As we mentioned before semi-NMF can be interpreted in terms of K-means clustering. For a given $E_l$, the learned basis matrix $M_l$ can be interpreted as matrix containing cluster centers. The weight matrix $W_i^l$ ($ n_i \times k$) obtained for $X_i$ using $M_l$ can then be interpreted as posterior probabilities for each frame in $X_i$ with respect to cluster centers in $M_l$. Hence, we first normalize each row of $W_i^l$ to sum to $1$, to convert them into probability space. Then, we obtain $k$ dimensional histogram representation for $X_i$ corresponding to $M_l$ as
\begin{equation}
\vec{h}_i^l = \frac{1}{n_i} \sum_{j=1}^{n_i} \vec{w}_t \,\,;\,\, \vec{w}_t=t^{th} \,\, row \,\, of \,\, W_i^l
\end{equation}
This is done for all $M_l$ and hence for each training recording we obtain a total of $L$, $k$ dimensional histograms represented by $\vec{h}_i^l$.
\subsubsection{GMM based characterization of $W_l$}
\label{ssec:basic}
We also propose another way of capturing distribution in $W_l$ where we actually fit a mixture model to it. For a given sound class $E_l$, we first collect $W_i^l$ for all $X_i$ in training data. We then train a Gaussian Mixture Model $\mathcal{G}^l$ on the accumulated weight vectors. Let this GMM be $\mathcal{G}^l = \{\lambda_g,N(\vec{\mu}_g, \Sigma_g), g = 1 \,\,to \,\,G^l\}$, where $\lambda_g^l$, $\vec{\mu}_g^l$ and $\Sigma_g^l$ are the mixture weight, mean and covariance parameters of the $g^{th}$ Gaussian in $\mathcal{G}^l$. Once $\mathcal{G}^l$ has been obtained, for any $W_i^l$ we compute probabilistic  posterior assignment of weight vectors $w_t$ in $W_i^l$ according to Eq \ref{eq:addms} ($ Pr(g | \vec{w}_{t})$). $\vec{w}_t$ are again the rows in $W_i^l$. These soft-assignments are added over all $t$ to obtain the total mass of weight vectors belonging to the $g^{th}$ Gaussian ($P(g)^l_i$, Eq \ref{eq:addms}). Normalization by $n_i$ is done to remove the effect of the duration of recordings.
\begin{equation}
\label{eq:addms}
\resizebox{0.9\columnwidth}{!}{$ Pr(g | \vec{w}_{t}) = \frac{\lambda_{g}^l N(\vec{w}_{t} ; \vec{\mu}_g^l, \Sigma_g^l)}{\sum\limits_{p=1}^G \lambda_p^l N(\vec{w}_{t} ; \vec{\mu}_p^l, \Sigma_p^l)}; P(g)^l_i = \frac{1}{n_i}\sum\limits_{t=1}^{n_i} Pr(g | \vec{w}_{t}) $}
\end{equation}
The final representation for $W_i^l$ is $\vec{v}_i^l=[P(1)^l_i,...P(G^l)^l_i]^T$. $\vec{v}_i^l$ is a $G^l$-dimensional feature representation for a given recording $X_i$ with respect to $E_l$. The whole process is done for all $E_l$ to obtain $L$ different soft assignment histograms for a given $X_i$. 
\subsection{Kernel Fusion for Semantic Content based Prediction}
\label{ssec:fsinfo}
$\vec{h}_i^l$ or $\vec{v}_i^l$ features captures acoustic events information for any $X_i$. We then use kernel fusion methods in Support Vector Machine (SVM) to finally train classifiers for geotagging purposes. We explain the method here in terms of $\vec{h}_i^l$, for $\vec{v}_i^l$ the steps followed are same. 

For each $l$, we obtain separate kernels representation $K_l$ using $\vec{h}_i^l$ for all $X_i$. Since exponential $\chi ^2$ kernel SVM are known to work well with histogram representations \cite{zhang2007local} \cite{cao2011ibm}, we use kernels of the form $K_l(\vec{h}_i^l,\vec{h}_j^l)=exp(-D(\vec{h}_i^l,\vec{h}_j^l)/\gamma)$ where $D(\vec{h}_i^l,\vec{h}_j^l))$ is $\chi^2$ distance between $\vec{h}_i^l$ and $\vec{h}_j^l$. $\gamma$ is set as the average of all pair wise distance. Once we have all $K_l$, we use two simple kernel fusion methods; 
\begin{itemize}
\item Average kernel fusion - The final kernel representation is given by, $K_S^h=\frac{1}{L} \sum_{l=1}^L K_l(;,;)$
\item Product kernel fusion - In this case the final kernel representation is given by, $K_P^h=\frac{1}{L} \prod_{i=1}^L K_l(:,:)$.
\end{itemize}
Finally, $K_S^h$ or $K_P^h$ is used to train SVMs for prediction.  
\section{Experiments and Results}
\label{sec:expt}
As stated before, our goal is to perform city - level geotagging in multimedia. Hence, we evaluate our proposed method on the dataset used in \cite{lei2012multimodal} which provides city level tags for flickr videos. The dataset contains contains a total of 1079 Flickr videos with $540$ videos in the training set and $539$ in the testing set. We work with only audio of each video and we will alternatively refer to these videos as audio recordings. The maximum duration of recordings is $90$ seconds. The videos of the recording belong to $18$ different cities with several cities having very few examples in training as well as testing set such as just $3$ for Bankok or $5$ for Beijing. We selected $10$ cities for evaluation for which training as well as test set contains at least $11$ examples.  These $10$ cities are \emph{Berlin (B), Chicago (C), London (L), Los Angeles (LA), Paris (P), Rio (R), San Francisco (SF), Seoul(SE), Sydney (SY)} and \emph{Tokyo (T)}. As stated before the basic acoustic feature used are MFCC features augmented by delta and acceleration coefficients. $20$ dimensional MFCCs are extracted for each audio recording over a window of $30$ $ms$ with $50\%$ overlap. Hence, basic acoustic features for audio recordings are $60$ dimensional and referred to as MFCA features.  

We compare our proposed method with two methods, one based on GMM based bag of audio words (BoAW) and other based on GMM-supervectors. These are clip level feature representation built over MFCA acoustic features for each recording. The first step in this method is to train a background GMM $\mathcal{G}^{bs}$ with $G^b$ components over MFCA features where each Gaussian represents an audio word. Then for each audio recording clip level histogram features are obtained using the GMM posteriors for each frame in the clip. The computation is similar to Eq \ref{eq:addms}; except that the process is done over MFCA features. These clip level representation are soft count bag of audio words representation. GMM-supervectors are obtained by adapating means of background GMM $\mathcal{G}^{bs}$ for a given using \emph{maximum a posteriori} (MAP) adaptation \cite{campbell2006support}. We will use $\vec{b}$ to denote these $G^b$ dimensional bag of audio words features and $\vec{s}$ to denote the $G^b \times 60$ dimensional GMM - supervectors. Exponential $\chi^2$ kernel SVMs are  used with $\vec{b}$ features and linear SVMs are used with GMM - supervectors features. Exponential $\chi^2$ kernels are usually represented as $K(x,y) = \exp^{-\gamma D(x,y)}$, where $D(x,y)$ is $\chi^2$ distance between vetors $x$ and $y$. Both of these kernels are known to work best for the corresponding features. All parameters such as $\gamma$ and the slack parameter $C$ in SVMs are selected by cross validation over the training set. 

For our proposed method we need a set of sound classes $E$. Studies on \emph{Urban soundscapes} have tried to categorize the urban acoustic environments \cite{brown2011towards} \cite{payne2009research} \cite{schafer1993soundscape}. \cite{salamon2014dataset} came up with a refined taxonomy of \emph{urban sounds} and also created a dataset, \emph{UrbanSounds8k}, for urban sound events. This dataset contains $8732$ audio recordings spread over $10$ different sound events from urban sound taxonomy. These sound events are car horn, children playing, dog barking, air conditioner noise, drilling, engine idling, gun shot, jackhammer, siren and street music. We use these $10$ sound classes as our set $E$ and then obtain the basis matrices $M_l$ for each $E_l$ using the examples of these sound events provided in the UrbanSounds8k dataset. 

The number of basis vectors for all $M_l$ is same and fixed to either $20$ or $40$. We present results for both cases. Finally, in the classifier training stage; SVMs are trained using the fused kernel $K_S^h$ (or $K_P^h$, or $K_S^v$, or $K_P^v$) as described in Section \ref{ssec:fsinfo}. Here the slack parameter $C$ in SVM formulation is set by performing $5$ fold cross validation over the training set. 

We formulate the geotagging problem as retrieval problem where the goal is to retrieve most relevant audios for a city. We use well known Average Precision (AP) as metric to measure performance for each city and Mean Average Precision (MAP) over all cities as the overall metric. Due to space constraints we are not able to show AP results in every case and will only present overall metric MAP. 
\begin{table}[t]
\centering
\caption{MAP for different cases ($\vec{b}$, $\vec{s}$ and $\vec{h}^l$) }
\resizebox*{1.0\columnwidth}{!}{
\begin{tabular}{|c|c|c|c|c|}
\hline  
$G^b$ $\rightarrow$ & 32 & 64 & 128 & 256 \\
\hline
MAP ($\vec{b}$) $\rightarrow$ & 0.362 & 0.429 & 0.461 & \textbf{0.478} \\
\hline
MAP ($\vec{s}$) $\rightarrow$ & 0.446 & \textbf{0.491} & 0.471 & 0.437 \\
\hline
\hline
\hline
Kernel $\rightarrow$ &\multicolumn{2}{c|}{Avg Ker. ($K_S^h$)} &  \multicolumn{2}{c|}{Prod. Ker ($K_P^h$)}\\ 
\hline
$k$ $\rightarrow$ & 20 & 40 & 20 & 40\\
\hline
MAP $\rightarrow$ & 0.454 & 0.500 & \textbf{0.520} & \textbf{0.563}\\
\hline
\end{tabular}
}
\label{tab:bspost}
\vspace{-0.15in}
\end{table}

Table \ref{tab:bspost} shows MAP results for BoAW and Supervector based methods (top $3$ rows) and our proposed method (bottom $3$ rows) using $\vec{h}^l$ features described in Section \ref{ssec:postr}. For baseline method we experimented with $4$ different component size $G^b$ for GMM $\mathcal{G}^{bs}$. $k$ represents the number of basis vectors in each $M_l$. $K_S^h$ represents the average kernel fusion and $K_P^h$ product kernel fusion. First, we observe that our proposed method outperforms these state of art methods by a significant margin. For BoAW, $G^b=256$ gives highest MAP of $0.478$ but MAP saturates with increasing $G^b$ and hence, any significant improvement in MAP by further increasing $G^b$ is not expected. For supervectors $G^b=64$ gives best result and MAP decreases on further increasing $G^b$. Our proposed method with $k=40$ and product kernel fusion gives $\mathbf{0.563}$ MAP, an absolute improvement of $\mathbf{8.5\%}$ and $\mathbf{7.2\%}$ when compared to BoAW and supervectors based methods respectively. MAP in other cases for our proposed method are also in general better than best MAP using state of art methods. We also note that for $\vec{h}^l$ features, product kernel fusion of different sound class kernels performs better than average kernel fusion. Also, for $\vec{h}^l$, $k=40$ is better than $k=20$. 

Table \ref{tab:basic} shows results for our $\vec{v}^l$ features in Section \ref{ssec:basic} which uses GMM based characterization of composition matrices $W_l$. We experimented with $4$ different values of GMM component size $G^l$. Once again we observe that overall this framework works gives superior performance. Once again MAP of $\mathbf{0.527}$ with $\vec{v}^l$ is over $3.6\%$ higher in absolute terms when comapred to best MAP with supervectors. 

This shows that the composition matrices $W_l$ are actually capturing semantic information from the audio and these semantic information when combined helps in location identification.  If we compare $\vec{v}^l$ and $\vec{h}^l$ methods then overall $\vec{h}^l$ seems to give better results. This is worth noting since it suggests that $W_l$ on its own are extremely meaningful and sufficient. Another interesting observation is that for $\vec{v}^l$ average kernel fusion is better than product kernel fusion. 

\begin{figure}[t] 
\centering
\includegraphics[trim=1.0in 0.50in 1.0in 0.7in,width=1.0\columnwidth,height=1.75in]{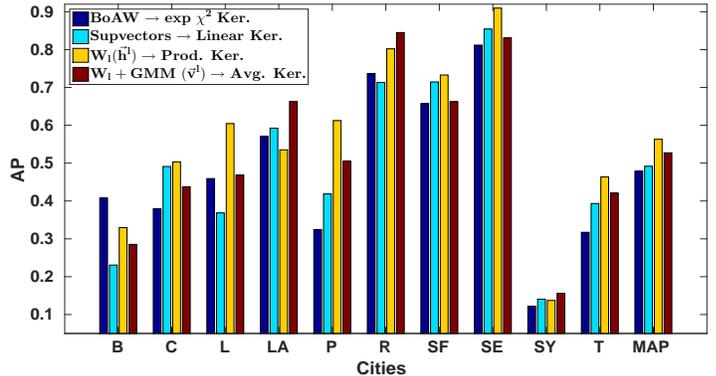}
\caption{Average Precision for Cities (MAP in right extreme)}
\label{fig:ctres}
\vspace{-0.10in}
\end{figure}

Figure \ref{fig:ctres} shows city wise results for all 4 methods methods. For each method the shown AP correspond to the case which results in best MAP for that method. This implies GMM component size in both BoAW and $\vec{v}^l$ is $256$ that is $G^l=G^b=256$; for $\vec{h}^l$ $k=40$ and  product kernel fusion; for $\vec{v}^l$ $k=20$ and average kernel fusion. For supervector based method $G^b=64$. For convenience, city names have been denoted by indices used in the beginning paragraph of this section. Figure \ref{fig:ctres} also shows MAP values in the extreme right. One can observe from Figure \ref{fig:ctres} that cities such as \emph{Rio (R), San Francisco (SF), Seoul (SE)} are much easier to identify and all methods give over $0.60$ AP. On the other hand \emph{Sydney (SY)} is a much harder to geotag comapred to other cities. Once again our proposed method outperforms BoAW and supervector based methods for all cities except for Berlin (B).    
\begin{table}[t]
\centering
\caption{MAP for different cases for $\vec{v}^l$ }
\resizebox*{1.0\columnwidth}{!}{
\begin{tabular}{|c|c|c|c|c|}
\hline  
&\multicolumn{2}{c|}{Avg Ker. ($K_S^v$)} &  \multicolumn{2}{c|}{Prod. Ker ($K_P^v$)}\\ 
\cline{2-5}
$G^l \downarrow$ \hspace{0.05in} | \hspace{0.05in} $k\rightarrow$ & 20 & 40 & 20 & 40\\
\hline
32 & 0.454 & 0.427 & 0.448 & 0.417\\
\hline
64 & 0.482 & 0.466 & 0.432 & 0.424\\
\hline
128 & \textbf{0.510} & 0.465 & 0.466 & 0.427\\
\hline
256 & \textbf{0.527} & 0.455 & 0.471 & 0.441\\
\hline
\end{tabular}
}
\label{tab:basic}
\end{table}
\vspace{-0.10in}
\section{Discussions and Conclusions}
\label{sec:cncls}
We presented methods for geotagging in multimedia using its audio content. We proposed that the semantic content of the audio captured in terms of different sound events which occur in our environment, can be used for location identification purposes. It is expected that larger the number of sound classes in $E$ the more distinguishing elements we can expect to obtain and the better it is for geotagging. Hence, it is desirable that any framework working under this idea should be scalable in terms of number of sounds in $E$. In our proposed framework the process of learning basis matrices $M_l$ are independent of each other and can be easily parallelized. Similarly, obtaining composition weight matrices $W_i^l$ can also be computed in parallel for each $E_l$ and so do the features $\vec{h}_i^l$ (or $\vec{v}_i^l$) and kernel matrices. Hence, our proposed is completely scalable in terms of number sound events in the set $E$. If required, one can also easily add any new sound class to an existing system if required. Moreover, our proposed framework can be applied on any acoustic feature. 

Even with $10$ sound events from urban sound taxonomy we obtained reasonably good performance. Our proposed framework outperformed state of art supervector and bag of audio word based methods by a significant margin. Currently, we used simple kernel fusion methods to combine event specific kernels. One can potentially use established methods such as multiple kernel learning at this step. This might lead to further improvement in results.  One can also look into other methods for obtaining basis matrices for sound events. A more comprehensive analysis on a larger dataset with larger number of cities can through more light on the effectiveness of the proposed method. However, this work does give sufficient evidence towards success of audio content based geotagging in multimedia.

\bibliographystyle{IEEEbib}
\bibliography{references}

\begin{thebibliography}{10}

\bibitem{luo2011geotagging}
Jiebo Luo, Dhiraj Joshi, Jie Yu, and Andrew Gallagher,
\newblock ``Geotagging in multimedia and computer vision—a survey,''
\newblock {\em Multimedia Tools and Applications}, vol. 51, no. 1, pp.
  187--211, 2011.

\bibitem{bao2012location}
Jie Bao, Yu~Zheng, and Mohamed~F Mokbel,
\newblock ``Location-based and preference-aware recommendation using sparse
  geo-social networking data,''
\newblock in {\em Proceedings of the 20th International Conference on Advances
  in Geographic Information Systems}. ACM, 2012, pp. 199--208.

\bibitem{bao2015recommendations}
Jie Bao, Yu~Zheng, David Wilkie, and Mohamed Mokbel,
\newblock ``Recommendations in location-based social networks: a survey,''
\newblock {\em GeoInformatica}, vol. 19, no. 3, pp. 525--565, 2015.

\bibitem{majid2013context}
Abdul Majid, Ling Chen, Gencai Chen, Hamid~Turab Mirza, Ibrar Hussain, and John
  Woodward,
\newblock ``A context-aware personalized travel recommendation system based on
  geotagged social media data mining,''
\newblock {\em International Journal of Geographical Information Science}, vol.
  27, no. 4, pp. 662--684, 2013.

\bibitem{medeval}
MediaEval,
\newblock ``\url{http://www.multimediaeval.org/},'' 2015.

\bibitem{choi2014placing}
J~Choi, B~Thomee, G~Friedland, L~Cao, K~Ni, D~Borth, B~Elizalde, L~Gottlieb,
  C~Carrano, R~Pearce, et~al.,
\newblock ``The placing task: A large-scale geo-estimation challenge for
  social-media videos and images,''
\newblock in {\em Proceedings of the 3rd ACM Multimedia Workshop on Geotagging
  and Its Applications in Multimedia}. ACM, 2014, pp. 27--31.

\bibitem{trevisiol2013retrieving}
Michele Trevisiol, Herv{\'e} J{\'e}gou, Jonathan Delhumeau, and Guillaume
  Gravier,
\newblock ``Retrieving geo-location of videos with a divide \& conquer
  hierarchical multimodal approach,''
\newblock in {\em Proceedings of the 3rd ACM conference on International
  conference on multimedia retrieval}. ACM, 2013, pp. 1--8.

\bibitem{song2012web}
Yi-Cheng Song, Yong-Dong Zhang, Juan Cao, Tian Xia, Wu~Liu, and Jin-Tao Li,
\newblock ``Web video geolocation by geotagged social resources,''
\newblock {\em Multimedia, IEEE Transactions on}, vol. 14, no. 2, pp. 456--470,
  2012.

\bibitem{kelm2013novel}
Pascal Kelm, Sebastian Schmiedeke, Jaeyoung Choi, Gerald Friedland,
  Venkatesan~Nallampatti Ekambaram, Kannan Ramchandran, and Thomas Sikora,
\newblock ``A novel fusion method for integrating multiple modalities and
  knowledge for multimodal location estimation,''
\newblock in {\em Proceedings of the 2nd ACM international workshop on
  Geotagging and its applications in multimedia}. ACM, 2013, pp. 7--12.

\bibitem{choi2013human}
Jaeyoung Choi, Howard Lei, Venkatesan Ekambaram, Pascal Kelm, Luke Gottlieb,
  Thomas Sikora, Kannan Ramchandran, and Gerald Friedland,
\newblock ``Human vs machine: establishing a human baseline for multimodal
  location estimation,''
\newblock in {\em Proceedings of the 21st ACM international conference on
  Multimedia}. ACM, 2013, pp. 867--876.

\bibitem{lei2012multimodal}
Howard Lei, Jaeyoung Choi, and Gerald Friedland,
\newblock ``Multimodal city-verification on flickr videos using acoustic and
  textual features,''
\newblock in {\em 2012 IEEE International Conference on Acoustics, Speech and
  Signal Processing (ICASSP)}. IEEE, 2012, pp. 2273--2276.

\bibitem{sevillano2012audio}
Xavier Sevillano, Xavier Valero, and Francesc Al{\'\i}as,
\newblock ``Audio and video cues for geo-tagging online videos in the absence
  of metadata,''
\newblock in {\em Content-Based Multimedia Indexing (CBMI), 2012 10th
  International Workshop on}. IEEE, 2012, pp. 1--6.

\bibitem{brown2011towards}
AL~Brown, Jian Kang, and Truls Gjestland,
\newblock ``Towards standardization in soundscape preference assessment,''
\newblock {\em Applied Acoustics}, vol. 72, no. 6, pp. 387--392, 2011.

\bibitem{payne2009research}
SR~Payne, WJ~Davies, and MD~Adams,
\newblock ``Research into the practical and policy applications of soundscape
  concepts and techniques in urban areas,''
\newblock 2009.

\bibitem{ding2010convex}
Chris Ding, Tao Li, and Michael~I Jordan,
\newblock ``Convex and semi-nonnegative matrix factorizations,''
\newblock {\em Pattern Analysis and Machine Intelligence, IEEE Transactions
  on}, vol. 32, no. 1, pp. 45--55, 2010.

\bibitem{lee2001algorithms}
Daniel~D Lee and H~Sebastian Seung,
\newblock ``Algorithms for non-negative matrix factorization,''
\newblock in {\em Advances in neural information processing systems}, 2001, pp.
  556--562.

\bibitem{zhang2007local}
Jianguo Zhang, Marcin Marsza{\l}ek, Svetlana Lazebnik, and Cordelia Schmid,
\newblock ``Local features and kernels for classification of texture and object
  categories: A comprehensive study,''
\newblock {\em International journal of computer vision}, vol. 73, no. 2, pp.
  213--238, 2007.

\bibitem{cao2011ibm}
Liangliang Cao, Shih-Fu Chang, Noel Codella, Courtenay Cotton, Dan Ellis,
  Leiguang Gong, Matthew Hill, Gang Hua, John Kender, Michele Merler, et~al.,
\newblock ``Ibm research and columbia university trecvid-2011 multimedia event
  detection (med) system,''
\newblock .

\bibitem{campbell2006support}
William~M Campbell, Douglas~E Sturim, and Douglas~A Reynolds,
\newblock ``Support vector machines using gmm supervectors for speaker
  verification,''
\newblock {\em IEEE signal processing letters}, vol. 13, no. 5, pp. 308--311,
  2006.

\bibitem{schafer1993soundscape}
R~Murray Schafer,
\newblock {\em The soundscape: Our sonic environment and the tuning of the
  world},
\newblock Inner Traditions/Bear \& Co, 1993.

\bibitem{salamon2014dataset}
Justin Salamon, Christopher Jacoby, and Juan~Pablo Bello,
\newblock ``A dataset and taxonomy for urban sound research,''
\newblock in {\em Proceedings of the ACM International Conference on
  Multimedia}. ACM, 2014, pp. 1041--1044.

\end{thebibliography}

\end{document}